\documentclass[12pt,preprint]{aastex}

\newcommand\spitzer{\textit{Spitzer}}

\newcommand{\ltsimeq}{\raisebox{-0.6ex}{$\,\stackrel
        {\raisebox{-.2ex}{$\textstyle <$}}{\sim}\,$}}
\newcommand{\gtsimeq}{\raisebox{-0.6ex}{$\,\stackrel
        {\raisebox{-.2ex}{$\textstyle >$}}{\sim}\,$}}

\slugcomment{ApJ Accepted 30Jan09}

\shorttitle{Water in comets}
\shortauthors{Bockel\'ee-Morvan et al.}

\begin{document}


\title{WATER IN COMETS 71P/CLARK AND C/2004 B1 (LINEAR) WITH {\it SPITZER}}


\author{
DOMINIQUE BOCKEL\'EE-MORVAN\altaffilmark{1},\\
CHARLES E. WOODWARD\altaffilmark{2},
MICHAEL S. KELLEY\altaffilmark{3}, \\
DIANE H. WOODEN\altaffilmark{4}
}


\altaffiltext{1}{LESIA, Observatoire de Paris, 5 place Jules Janssen,
F92195, Meudon, France, \\ {\it{dominique.bockelee@obspm.fr}} }

\altaffiltext{2}{Department of Astronomy, School of Physics and
Astronomy, 116 Church Street, S.~E., University of Minnesota,
Minneapolis, MN 55455,\ {\it{chelsea@astro.umn.edu}} }

\altaffiltext{3}{Department of Physics, University
of Central Florida, 4000 Central Florida Blvd., Orlando, FL
32816-2385, \ {\it{msk@physics.ucf.edu}} }

\altaffiltext{4}{NASA Ames Research Center, MS~245-3, Moffet Field, CA
94035-1000, \ {\it{diane.h.wooden@nasa.gov}} }

\begin{abstract}

We present 5.5--7.6~\micron{} spectra of comets 71P/Clark
(2006 May 27.56~UT, $r_{h} = 1.57$~AU pre-perihelion) and
C/2004~B1 (LINEAR) (2005 October 15.22~UT, $r_{h} = 2.21$~AU pre-perihelion
and 2006 May 16.22~UT, $r_{h} = 2.06$~AU post-perihelion) obtained
with the \textit{Spitzer Space Telescope.} The $\nu_{2}$ vibrational band
of water is detected with a signal-to-noise ratio of 11 to 50. Fitting
the spectra using a fluorescence model of water emission yields a water
rotational temperature of $<$ 18~K for 71P/Clark and $\simeq 14 \pm 2$~K
(pre-perihelion) and $23 \pm 4$~K (post-perihelion) for C/2004~B1 (LINEAR).
The water ortho-to-para ratio in C/2004~B1 (LINEAR) is
measured to be $2.31 \pm 0.18$, which corresponds to a spin
temperature of $26^{+3}_{-2}$~K. Water production rates are derived.
The agreement between the water model and the measurements is good,
as previously found for \spitzer~spectra of C/2003 K4 (LINEAR).
The \textit{Spitzer} spectra of these three comets do not show any
evidence for emission from PAHs and carbonate minerals, in contrast
to results reported for comets 9P/Tempel~1 and C/1995 O1 (Hale-Bopp).

\end{abstract}



\section{INTRODUCTION\label{intro}}

Production rates of volatiles are used to quantify cometary activity and
to measure volatile abundances. Water is the dominant ice in cometary
nuclei and its sublimation governs the activity of comets at
heliocentric distances, $r_{h}$, $\ltsimeq 3$~AU from the Sun. Water
production rates, $Q$(H$_{2}$O), the rotational temperature, $T_{rot}$(K),
and the nuclear spin temperature inferred from the water ortho-to-para
ratio (OPR) are physical parameters useful in a variety of studies
related to cometary atmospheres and cometary physics.

Unlike other constituents of cometary atmospheres, water vapor is
difficult to observe directly from the ground because of
terrestrial atmospheric absorption. Non-resonance fluorescence
vibrational bands (hot bands), not absorbed by telluric H$_{2}$O,
can be observed from the ground in relatively active comets
\citep[$Q$(H$_{2}$O) $\gtsimeq$ a few 10$^{28}$
molecules~s$^{-1}$; ][]{dis08}. The detection of the H$_{2}$O
fundamental bands of vibration ($\nu_{1}$, $\nu_{3}$ at
2.7~$\mu$m, $\nu_{2}$ at 6.3~$\mu$m), and of H$_{2}$O rotational
lines in the submillimeter and far-infrared domains, requires
space-based instrumentation. Detection of these signatures has
been achieved in the coma of several comets using spectrographs on
the \textit{VEGA, Deep Impact,} and \textit{Rosetta} spacecrafts,
and instruments on the \textit{Kuiper Airborne Observatory}, the
\textit{Infrared Space Observatory (ISO)}, the
\textit{Submillimeter Wave Astronomy Satellite,} and the
\textit{Odin} satellite (\citealt{ahearn05,biv07,boc04} and
references therein; \citealt{gul07}).

The Infrared Spectrograph \citep[IRS;][]{houck04} of the \textit{Spitzer
Space Telescope} \citep{gehrz07, werner04} covers the
5.3--40~\micron{} spectral region and enables
observations of the $\nu_{2}$ vibrational band of water vapor at
6.3~$\mu$m with a spectral resolution of $\sim$100. This band was
detected with \textit{Spitzer} in comet C/2003 K4 (LINEAR) with a
high signal-to-noise ratio ($\gtsimeq 50$), which enabled a detailed
study of the band shape using a fluorescence excitation model, assessment
of the radial gradient of emission with cometocentric radius, and the
measurement of the rotational temperature and ortho-to-para ratio (OPR)
of the detected water \citep{woodw07}.

An important by-product of the analysis of the \spitzer{} spectra of
comet C/2003 K4 (LINEAR) performed by \citet{woodw07} was a search for
emissions in excess to the water band in the 5.8--7.6~\micron{} domain,
which is rich in spectral signatures of gaseous compounds and minerals.
Indeed, \citet{lisse06,lisse07} reported emission from carbonate minerals
at 6.5--7.2~\micron{} and organic (PAH) emission at 6.2~\micron{} in the
\spitzer{} spectrum of comet 9P/Tempel after the collision with the
\textit{Deep~Impact} impactor, and in comet C/1995 O1 (Hale-Bopp) based
on their re-analysis of the \citet{cro97b} {\it ISO} spectrum. The
methodology of \citet{lisse06,lisse07} does not rely on a search of
individual signatures but from the decrease of the residual $\chi^{2}$
after a fit of the whole (5.2 to 40~\micron{} for \spitzer{} spectra)
observed spectrum by a modeled spectrum which can incorporate many
constituents. A reanalysis of the {\it ISO} data of comet Hale-Bopp by
\citet{cro08} did not confirm the  detection of PAHs reported by
\citet{lisse06}, and suggests that carbonate signatures are at most
marginal. After detailed modeling of the water $\nu_{2}$ emission, the
spectra of C/2003 K4 (LINEAR) do not show any evidence for emission from
PAHs and carbonate minerals \citep{woodw07}. Carbonates have been
identified in very tiny quantities in the dust grains collected in the
coma of 81P/Wild 2 by the {\it Stardust}  spacecraft \citep{wir07}. PAHs
have been also detected in {\it Stardust}{} samples \citep{sand06}, but a
substantial portion of this material may not be of cometary origin
\citep{sand07,spen07}. PAHs were tentatively identified in near-IR and
near-UV cometary spectra, but these detections are controversial
\citep[see the discussion in][]{boc04}.

Here, we present longslit \spitzer{} spectroscopic observations of comets
71P/Clark and C/2004 B1~(LINEAR). Observations in the 5.3--7.6~\micron{}
spectral region are presented in \S~\ref{sec:obs}. The data are
analyzed with a H$_{2}$O fluorescence model using the same methodology
described in \citet{woodw07} and the results are presented in
\S~\ref{sec:fit}. Residual emissions are discussed in
\S~\ref{sec:residuals}. We conclude in \S~\ref{sec:conclusion}.

\section{OBSERVATIONS AND REDUCTION}
\label{sec:obs}

Comet 71P/Clark was discovered on 1973 June 09 by Michael Clark at
Mount John University Observatory in New Zealand. This comet,
with an orbital period of 5.52 years, belongs to the dynamical class
of Jupiter-family comets. Its last perihelion approach was on 2006
June 06.80 (perihelion distance $q$ = 1.562 AU). Little
information is available on this comet in the literature. The
radius of its nucleus has been estimated to $0.68 \pm 0.07$~km
from observations with the Hubble Space Telescope obtained at
$r_{h} = 2.62$~AU, to 1.3~km from observations at $r_{h} =
4.67$~AU performed with the Keck telescope, suggesting an
elongated nucleus with an axis ratio $a/b~\gtsimeq~2.85$
\citep{lam04,meech04}. The orbit of this comet is peculiar. 71P
has a non-zero, non-gravitational thrust in the direction normal
to the orbital plane (the A$_{3}$ non-gravitational parameter)
that is well-determined with  good confidence
\cite[e.g.,][]{Sekanina93,nakano06a,nakano06b}. Erratic time
variability of the non-gravitational parameters over nucleus
revolutions were interpreted as due to a change in the surface
distribution of active areas and of the orientation of the spin
axis \citep{Sekanina93,Szu99}. The water production rate of
71P/Clark at perihelion in 1973 and 1984 was estimated to be 8--9
$\times10^{27}$ molecules~s$^{-1}$ based on its visual brightness
\citep{Sekanina93}. \citet{woodw08} find strong evidence for CO
and CO$_{2}$ gaseous emission near 4.5~\micron{} in the extended
coma pre-perihelion ($r_{h} = 1.607$~AU) during the 2006
apparition. The \spitzer{} IRS observations of 71P/Clark were
conducted $\approx 10$~days before its 2006 perihelion passage.

Comet C/2004~B1 (LINEAR), first identified as an asteroidal object
with a visual magnitude of 19.1 on 2004 Jan 29.16~UT at a $r_{h}
\simeq 8.1$~AU, was discovered by \citet{iauc8279} to be a comet
on 2004 Jan 30.1~UT with a 3\arcsec{} coma, and an extended tail
($\approx 4$\arcsec{}, PA $\simeq 310$\degr). Orbital elements
derived from astrometric observations \citep{mpec2004-b73} and
subsequently refined over a data arc spanning years 2004 through
2007 (1779 observations), indicate that this comet is dynamically
new, highly inclined to the ecliptic ($i \simeq 114$\degr), with a
perihelion of 2006 Feb 07.89~UT with a periapsis, $q = 1.602$~AU.
Although a relatively bright comet, few ground-based observations
are currently reported in the literature. \citet{tbonev07}
conducted $R-$band imaging observations of C/2004~B1 (LINEAR) over
a $\sim60$ day period between 2006 June and 2006 August, reporting
that the comet underwent episodes of variable activity on
timescales of days, with a dust production rate \citep{ahearn84},
$Af\rho \simeq 200 - 350$~cm. The comet also exhibited a
substantial coma ($\approx 2\times10^{4}$~km diameter) near the
end of 2006 June, and slowly faded  in $R-$brightness, from
$\simeq 14.3$~mag to $\simeq 15.1$~mag, 193 days after perihelion.

\subsection{\textit{Spitzer IRS}\label{obs-irs}}

Comet 71P/Clark was observed once ($r_{h} = 1.565$~AU, a
\spitzer{}-comet distance of 0.920~AU, and a phase angle of 37.8\degr) in
the second order of the short-wavelength, low-resolution module (SL2)
on 2006 May 27.56 UT as part of a large \spitzer{} Cycle 2 comet survey,
program identification (PID) 20021 (PI: C.E. Woodward),
astronomical observation request (AOR) key
\dataset[ADS/Sa.Spitzer#13818368]{13818368}, and processed with
IRS reduction pipeline S15.3.0.  The SL2 slit is 3.7\arcsec{} wide
and provides 57\arcsec{} of spatially resolved spectra
(1.85\arcsec~pixel$^{-1}$) with a spectral dispersion of 0.06~\micron.
Nine spectra (60~s $\times$ 5 cycles) at 5.2--7.6~\micron{} were
recorded in a $3\times3$ spectral map (executed with no peak-up), with
$2.47\arcsec\times38.0\arcsec$ \ steps (perpendicular $\times$ parallel to
the long slit dimension). The comet was also observed with the
long-wavelength, low-resolution (LL) module of IRS, with which we used to
verify the pointing of the spacecraft. Further analysis of the LL
spectrum is beyond the scope of this paper.

Comet C/2004 B1 (LINEAR) was observed twice, both pre- and
post-perihelion with the SL2 module as part of a multi-cycle
\spitzer{} initiative to investigate the heliocentric evolution
and activity of a given comet, PID 20104 (PI: D.H. Wooden). The
SL2 spectra discussed here were derived from IRS spectral mapping
observations obtained on 2005 October 15.22~UT (pre-perihelion,
$r_{h} = 2.210$~AU) AOR key
\dataset[ADS/Sa.Spitzer#15787008]{15787008}, and on 2006 May
16.24~UT (post-perihelion, $r_{h} = 2.058$~AU), AOR key
\dataset[ADS/Sa.Spitzer#15791616]{15791616}. Pre-perihelion, three
spectra (240~s $\times$ 5 cycles) were obtained in from a
$3\times1$ spectral map, while post-perihelion five
spectra (60~s $\times$ 3 cycles) were
generated from a $5\times1$ spectral map. Both AORs were executed
using a mapping footprint of  2.50\arcsec{} steps (perpendicular
to the long slit dimension) with no peak-up on the comet nucleus,
accompanied by appropriate shadow (background) observations.
Spectral observations of comet C/2004 B1 (LINEAR) using the
short-high (SH, 9.9--19.6~\micron) module were also obtained, although
detailed analysis of these spectra are beyond the scope of this manuscript.

\begin{figure}
\begin{center}
\includegraphics[angle=0,scale=.90]{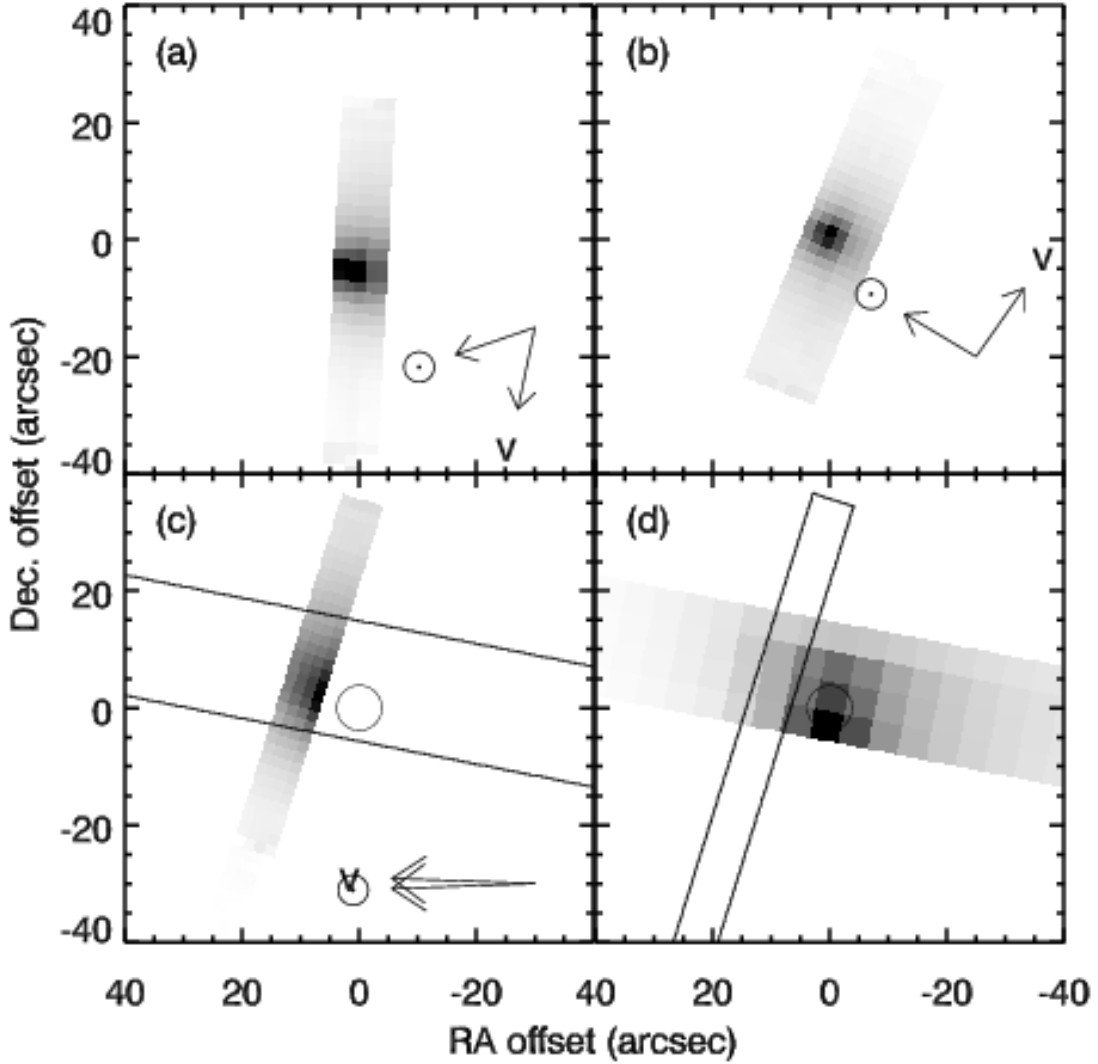}
\end{center}
\caption{Gray-scale surface-brightness images derived from
\spitzer{} IRS spectral mapping data cubes of comets C/2004 B1 (LINEAR) and
71P/Clark: (a) IRS short-low order 2 (SL2; 5.3--7.5~\micron) observation
of comet C/2004 B1 (LINEAR) at $r_{h}=2.2$~AU (pre-perihelion); (b) IRS
SL2 observation of comet C/2004 B1 (LINEAR) at $r_{h}=2.1$~AU
(post-perihelion); (c) IRS SL2 observation of comet 71P/Clark at
$r_{h}=1.6$~AU (pre-perihelion); (d) IRS long-low order 2
(LL2; 14--20~\micron) observation of comet
71P/Clark obtained as part of the same \spitzer{} AOR used to acquire the
SL2 observation of panel (c). Arrows in panels (a), (b), and (c) delineate
the projected sun ($\sun$) and comet velocity (v) directions. The 71P/Clark
SL2 observation did not cover the comet nucleus; however, the LL2
observation did.  To emphasize this
point, we have marked the JPL/Horizons ephemeris position of comet
71P/Clark with a circle, and placed outlines of the LL2 and SL2 modules
in panels (c) and (d), respectively. Spectral extractions discussed in
\S~\ref{sec:obs} are centered on the peak of the surface brightness
distribution.
\label{fig:pointing}}
\end{figure}

A reconstruction of the pointing for the 71P/Clark SL2 spectral
maps indicates that there is an 8\arcsec{} perpendicular offset
between the SL2 extraction position and the JPL
Horizons\footnote{http://ssd.jpl.nasa.gov/horizons.cgi} ephemeris
position of the comet nucleus at the epoch of our observations,
Fig.~\ref{fig:pointing}(c). This offset is confirmed by
comparing the peak in the observed surface brightness distribution
in the long-low (LL2, 14--20~\micron) cube which is nearly
coincident with the Horizons position, Fig.~\ref{fig:pointing}(d).
The spectral maps for both IRS pointings of C/2004~B1 (LINEAR) encompassed
the peak in the surface brightness of the coma (Fig.~\ref{fig:pointing}),
hence SL2 extraction apertures centered on the nucleus are possible.

The reduction of the spectra and wavelength calibration of the SL2
data follows the formalism discussed in detail by \cite{kelley06}
and \citet{woodw07}. ``Hot-pixels'' occasionally present in
individual extractions derived from the spectral maps were removed and
replaced by nearest neighbor interpolation, or ignored altogether. A
summary of comet physical parameters, spectral extraction apertures,
and derived water production rates (see \S~\ref{sec:fit}) are given
in Table~\ref{table:fit}.

\section{DATA ANALYSIS}
\label{sec:fit}

\subsection{\textit{Model Fitting of the $\nu_{2}$ Water Band}}
\label{sec:h2ofit}

\begin{figure}
\begin{center}
\includegraphics[angle=270,scale=.90]{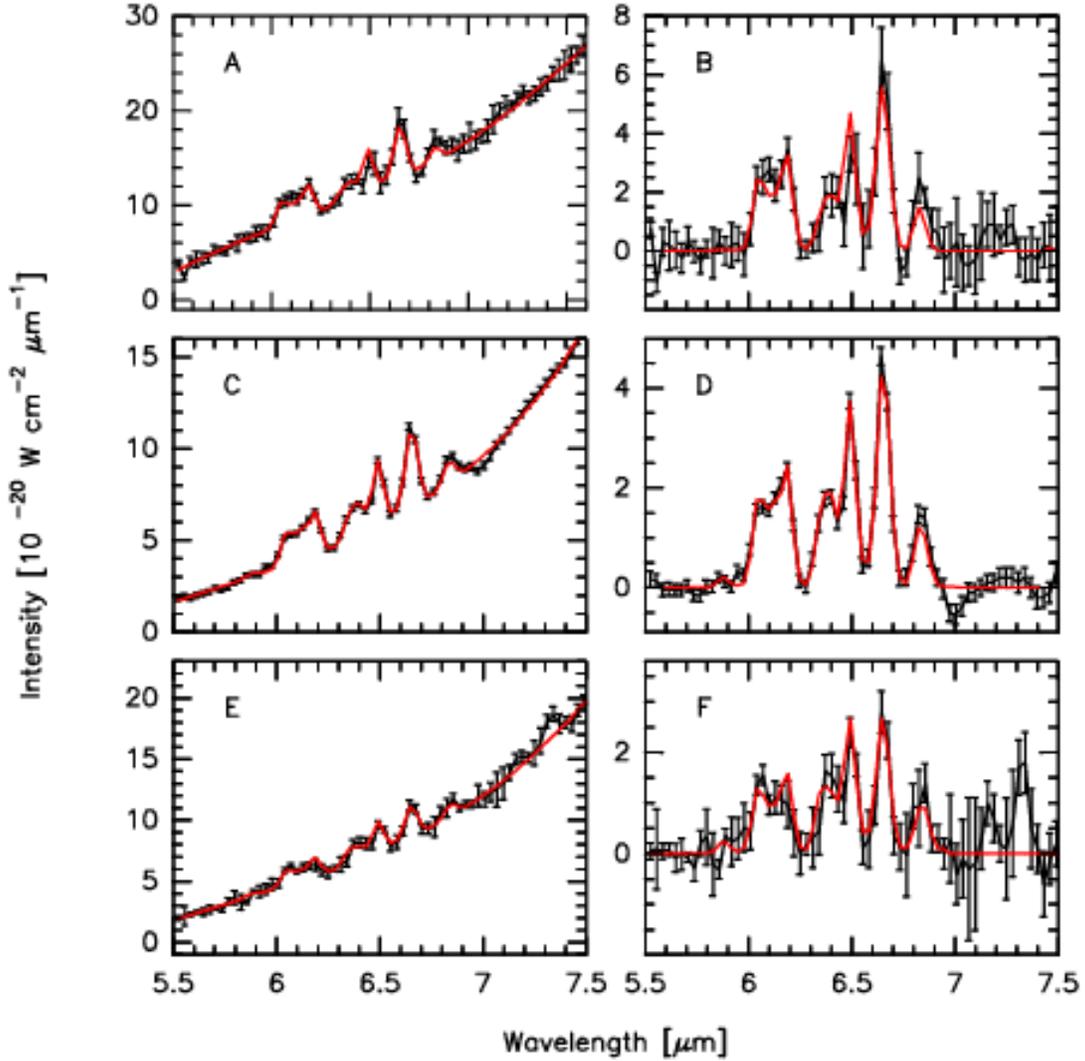}
\end{center}
\caption{Model fits to the spectra of 71P/Clark (A, B) and C/2004
B1 (LINEAR) pre-perihelion (C, D) and post-perihelion (E, F)
observed with \textit{Spitzer} on 2006 May 27.56~UT,  2005 October
27.56~UT and 2006 May 16.22~UT, respectively. The extraction
aperture is 3.7\arcsec $\times$ 17.9\arcsec. Data are shown
in black, with the model fits of water emission (in red)
superimposed. For 71P/Clark and C/2004 B1 (LINEAR)
post-perihelion, model fitting was performed with the rotational
temperature  $T_{rot}$ taken as a free parameter, and the
ortho-to-para ratio set to OPR = 3. In contrast, for the
pre-perihelion spectrum of C/2004 B1 (LINEAR), the fit was
obtained with both the OPR and $T_{rot}$ taken as a free parameters.
The underlying continuum, described by a polynomial of degree 5, was
also fitted simultaneously. In panels B and D, the fitted continuum
background has been subtracted. \label{fig:spectra}}
\end{figure}

As for C/2003 K4 (LINEAR), features observed between 5.8 and 7~\micron{}
above the continuum background in the 71P/Clark and C/2004~B1
(LINEAR) \spitzer~spectra (Fig.~\ref{fig:spectra}) coincide with
the ro-vibrational structure of the $\nu_{2}$ water band at the resolution
of the short-low (SL2) spectrometer. In order to study whether water band
emission fully accounts for the signal in excess of the continuum, these
spectra were analyzed using a model of fluorescence water emission 
following the procedure described in \citet {woodw07}. Details on 
the method, model and assumptions can be find in \citet{woodw07}.

In summary, we use the model of fluorescence water emission
presented by \citet{domi89} which considers five excited
vibrational states and their subsequent radiative cascades.
Indeed, the $\nu_{2}$ band, which dominates the water spectrum in
the 5.8--7.1~\micron{} spectral region, is significantly populated
by radiative decay from higher excited vibrational states. The
resulting emission rate of $\nu_{2}$ is $2.41 \times
10^{-4}$~s$^{-1}$ at $r_{h} = 1$~AU from the Sun. For computing
the line-by-line fluorescence, 32 ortho and 32 para rotational
levels are considered in each vibrational state. The rotational
populations in the ground vibrational state is described by a
Boltzmann distribution at a temperature $T_{rot}$. The gas
expansion velocity, $v_{exp}$, was fixed to 0.8~km~s$^{-1}$ and
the water photodissociation rate was taken equal to $1.6 \times
10^{-5}$~s$^{-1}$ ($r_{h} = 1$~AU). These parameters influence to
a small extent the retrieved water production rates.


From the model output, synthetic \spitzer{} spectra are
generated by convolving the intensity of the individual
ro-vibrational lines with the instrumental spectral response of
the spectrometer. The spectral resolution is determined to be
$0.0614 \pm 0.0027$~\micron{} from the pre-perihelion spectrum of
C/2004 B1 (LINEAR), in good agreement with the SL2 spectral
resolution (0.0605~\micron) cited in the \spitzer{} IRS Data Hand 
Book v.3.2\footnote{\url{http://ssc.spitzer.caltech.edu/IRS/dh}}. 
Our derived value is used for the analysis of the comet spectra in 
this paper. A slightly lower value of the SL2 spectral resolution, 
0.065~\micron{} (determined from the wavelength calibration at the SL2 
slit edge), was used in the analysis of the C/2003 K4 (LINEAR) spectra 
as spectral extractions of this comet were performed near the edge
of the slit \citep[see \S~2.2 and 3.1 in][]{woodw07}.


The observed spectra are fitted with a composite curve consisting
of the modeled water spectrum superimposed on a 5$^{th}$--degree
polynomial. This latter polynomial determines the underlying
continuum due to dust emission. The best-fit modeled spectrum is
searched for by applying a least-squares method that uses the
gradient-search algorithm of Marquardt. The only free parameters
of the model are the water production rate $Q$(H$_{2}$O) (or
equivalently, the water column density), $T_{rot}$, the
ortho-to-para ratio (OPR), and the polynomial for the continuum.
However, the signal-to-noise ratio in the spectra of comets
71P/Clark and C/2004~B1 (LINEAR) post-perihelion is not high
enough to constrain the OPR, so this parameter was fixed when
analyzing these data.

Figure~\ref{fig:spectra} shows the model fits superimposed on the
observed spectra. In the right panels, B, D and F, the continuum
has been subtracted. Model fits with residuals are shown in
Figs.~\ref{fig:71pfit} and ~\ref{fig:post-2004b1fit}, for
71P/Clark and C/2004 B1 (LINEAR), respectively. The intensity of
the water band in continuum-subtracted spectra and the retrieved
$T_{rot}$, OPR and $Q$(H$_2$O) are given in Table~\ref{table:fit}.

\begin{figure}
\begin{center}
\includegraphics[angle=270,scale=.50]{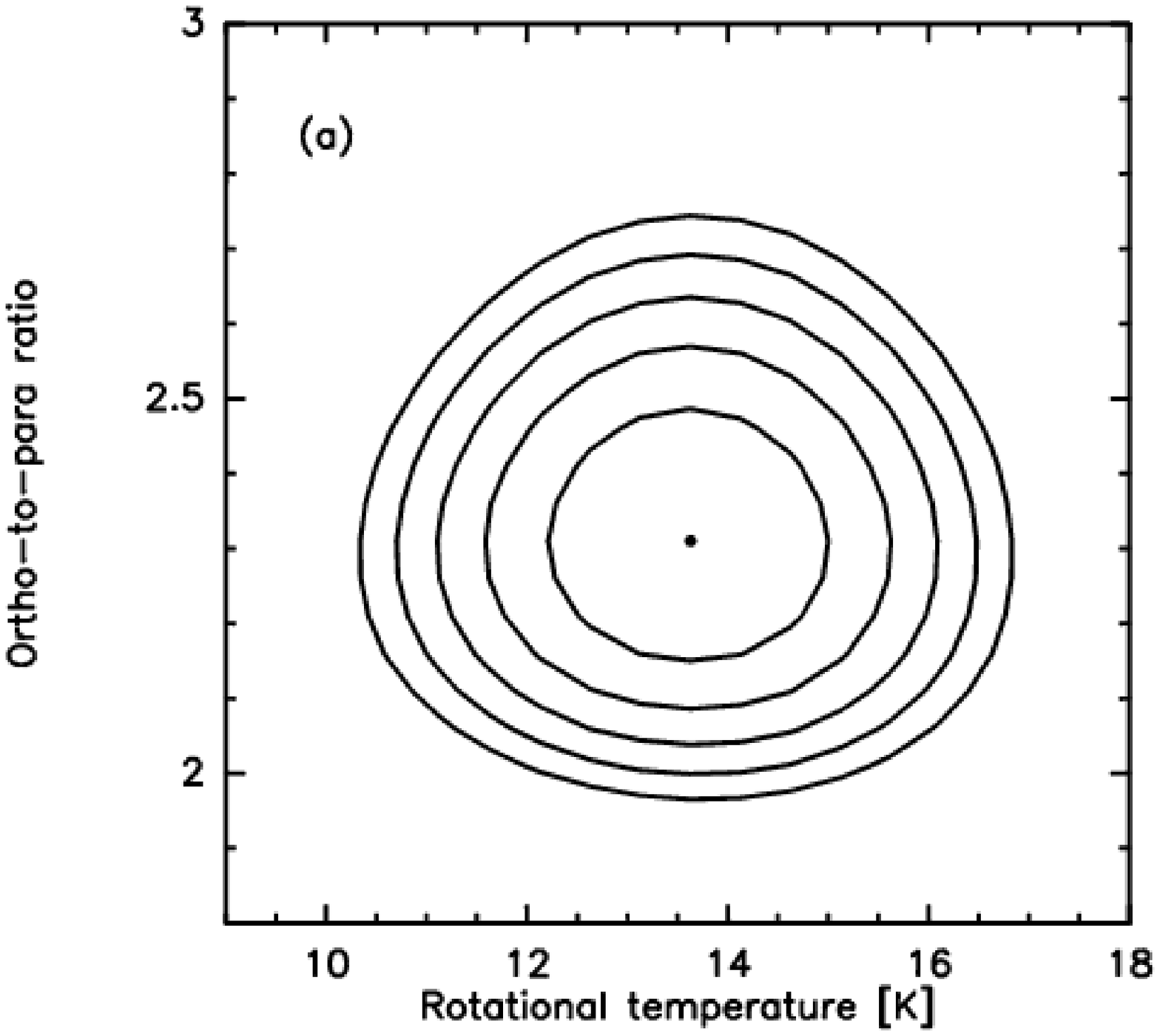}
\includegraphics[angle=270,scale=.50]{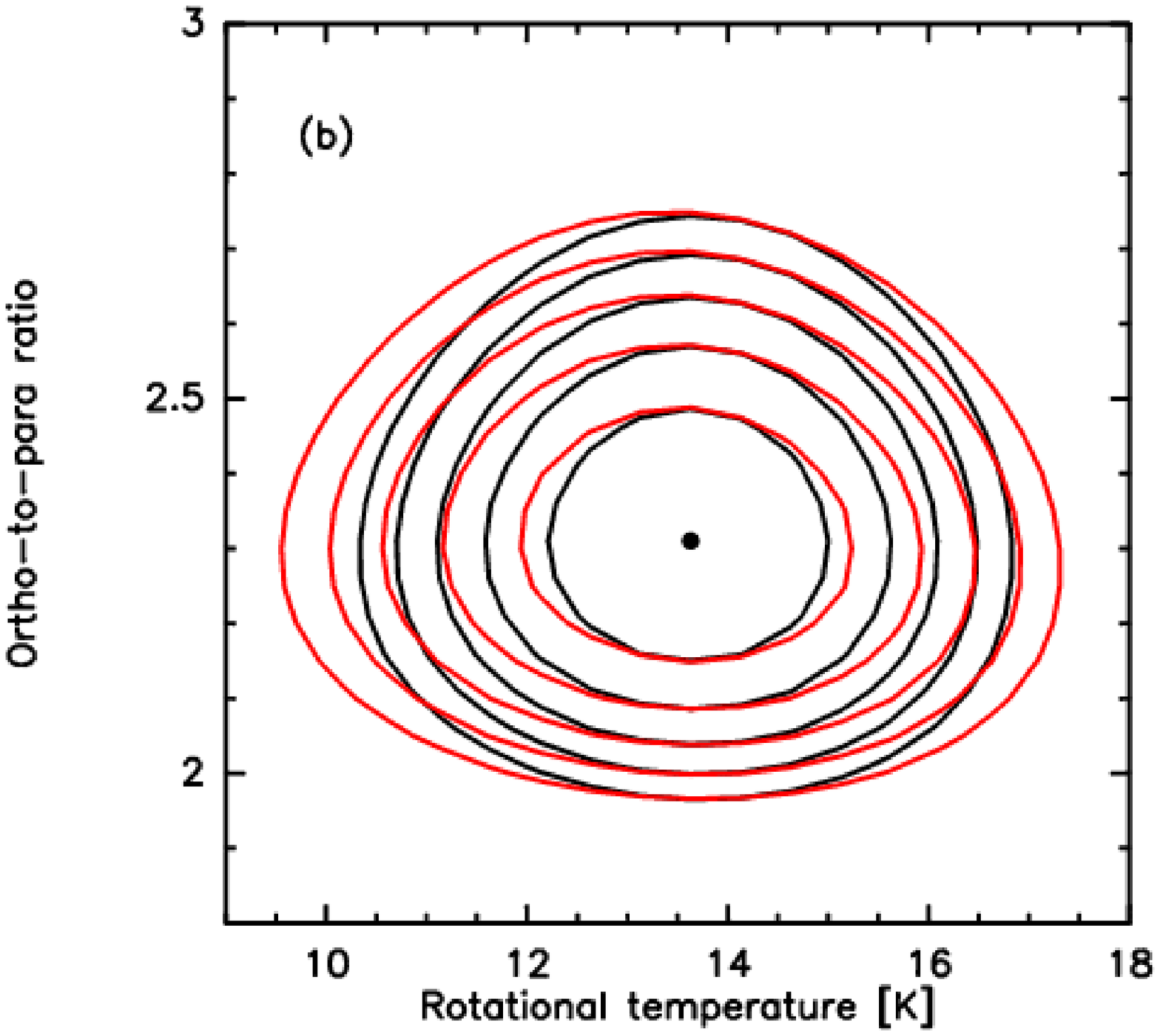}
\end{center}
\vspace{-0.75cm}
\caption{Contours showing the variation of $\chi^2$ with
$T_{rot}$ and OPR in the region that provides the best fit
to the C/2004 B1 (LINEAR) pre-perihelion spectrum. Iso-$\chi^2$ contour
levels correspond to increases in $\chi^2$ by steps of 1 from the minimum
value (denoted by small black dot). (a)~The contours are 
calculated by holding the polynomial
coefficients and the water production rate fixed at their optimum values.
(b)~The contours shown in red color are calculated by allowing
those parameters to vary to minimize $\chi^2$ for each pair of values of
$T_{rot}$ and OPR. Black contours are those shown in plot (a).
\label{fig:chi2-map}}
\end{figure}

As discussed in \citet{woodw07}, the spectral resolution of
\spitzer~SL2 is to low to resolve the individual ortho and para
water lines. Thus a fully independent estimate of the OPR and
$T_{rot}$ cannot be derived. However, a robust determination 
of these parameters is possible from model fitting and 
$\chi^2$ minimization, because
several para lines (near 6.12 and 6.4~\micron) are well separated
in wavelength from strong ortho lines, while several peaks,
dominated by ortho lines, have their relative intensities
essentially influenced by $T_{rot}$. Spectra obtained with high
signal-to-noise ratios are of course requisite.
Figure~\ref{fig:chi2-map}(a) shows iso-$\chi^2$ contours in the
($T_{rot}$, OPR) region that provides the best fit to the 
C/2004~B1 (LINEAR) pre-perihelion spectrum. The coefficients of the
polynomial describing the continuum background have been set to
the values of the best fit. The reduced $\chi^2$ obtained for the
best fit is 0.97 in the 5.85--7~\micron{} region. The weak
inclination of the ellipses (i.e., their roundness at the scales
of the plot) indicates that the two parameters are poorly
correlated \citep{bevington}. The linear correlation coefficient 
is equal to 0.034.  In Figure~\ref{fig:chi2-map}(b), red iso-$\chi^2$ 
contours are computed by fixing $T_{rot}$ and OPR and then minimizing 
$\chi^2$ with the coefficients of the polynomial and the water production
rate left as free parameters. With respect to the iso-$\chi^2$ contours
computed with the previous method (superimposed in black), these
$\chi^2$ iso-contours extend a bit farther in $T_{rot}$, while
they coincide along the OPR dimension. This demonstrates that
$T_{rot}$ is weakly correlated with the coefficients of the
polynomial, whereas OPR is not correlated. Quoted uncertainties in
Table~\ref{table:fit} and throughout the paper are 1-$\sigma$
(68\% confidence level) uncertainties computed from the diagonal
elements of the covariance matrix. They correspond approximatively
to the $\Delta\chi^2$ = 1 red contour in
Figure~\ref{fig:chi2-map}(b). The joint confidence region for
($T_{rot}$, OPR), delimited by the $\Delta\chi^2$ = 2.3 contour 
\citep{bevington}, is not used here as these parameters are not 
physically linked.


\subsection{\textit{Water Modeling Results}}

\begin{figure}
\begin{center}
\includegraphics[angle=270,scale=.40]{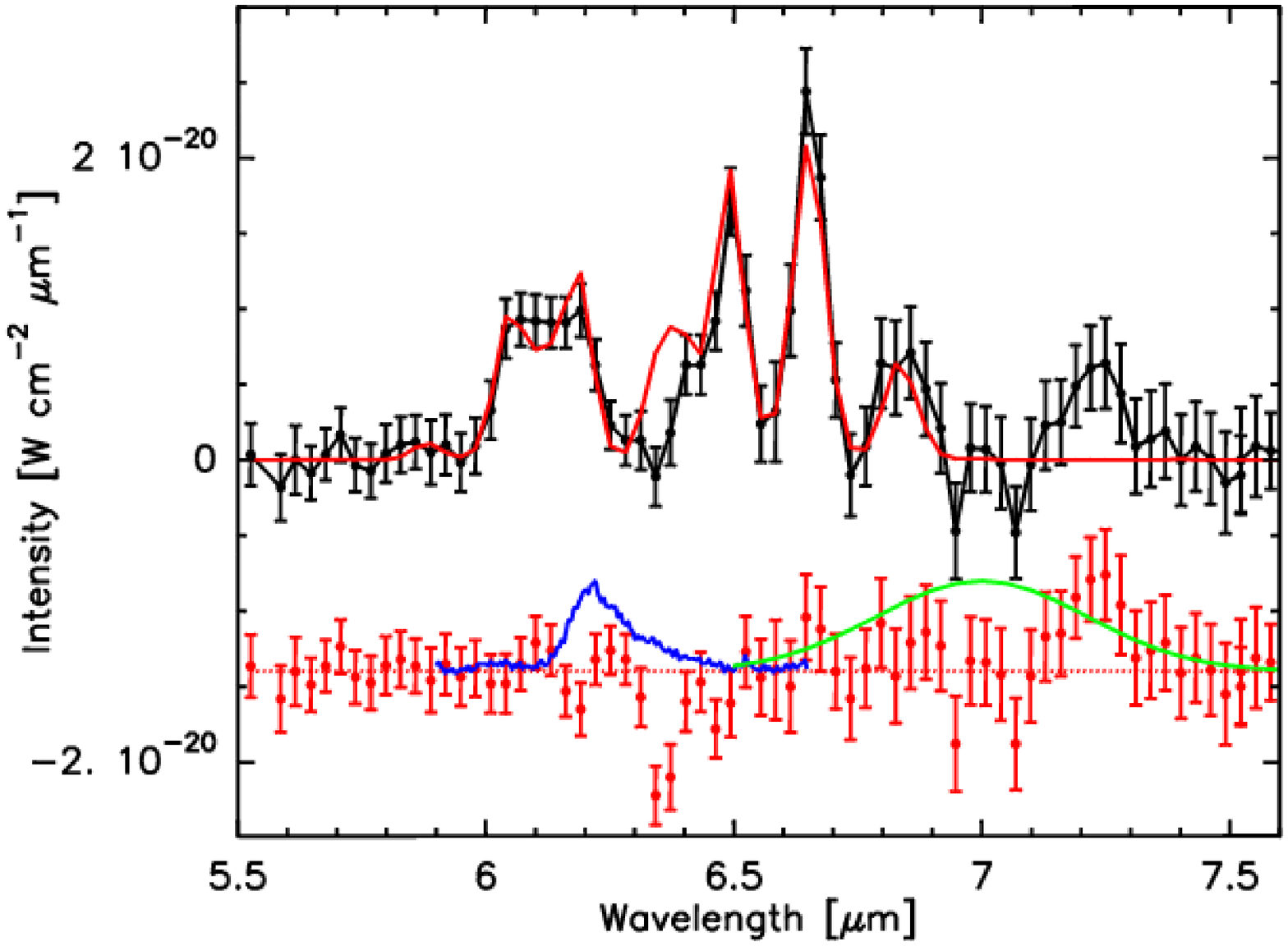}
\includegraphics[angle=270,scale=.40]{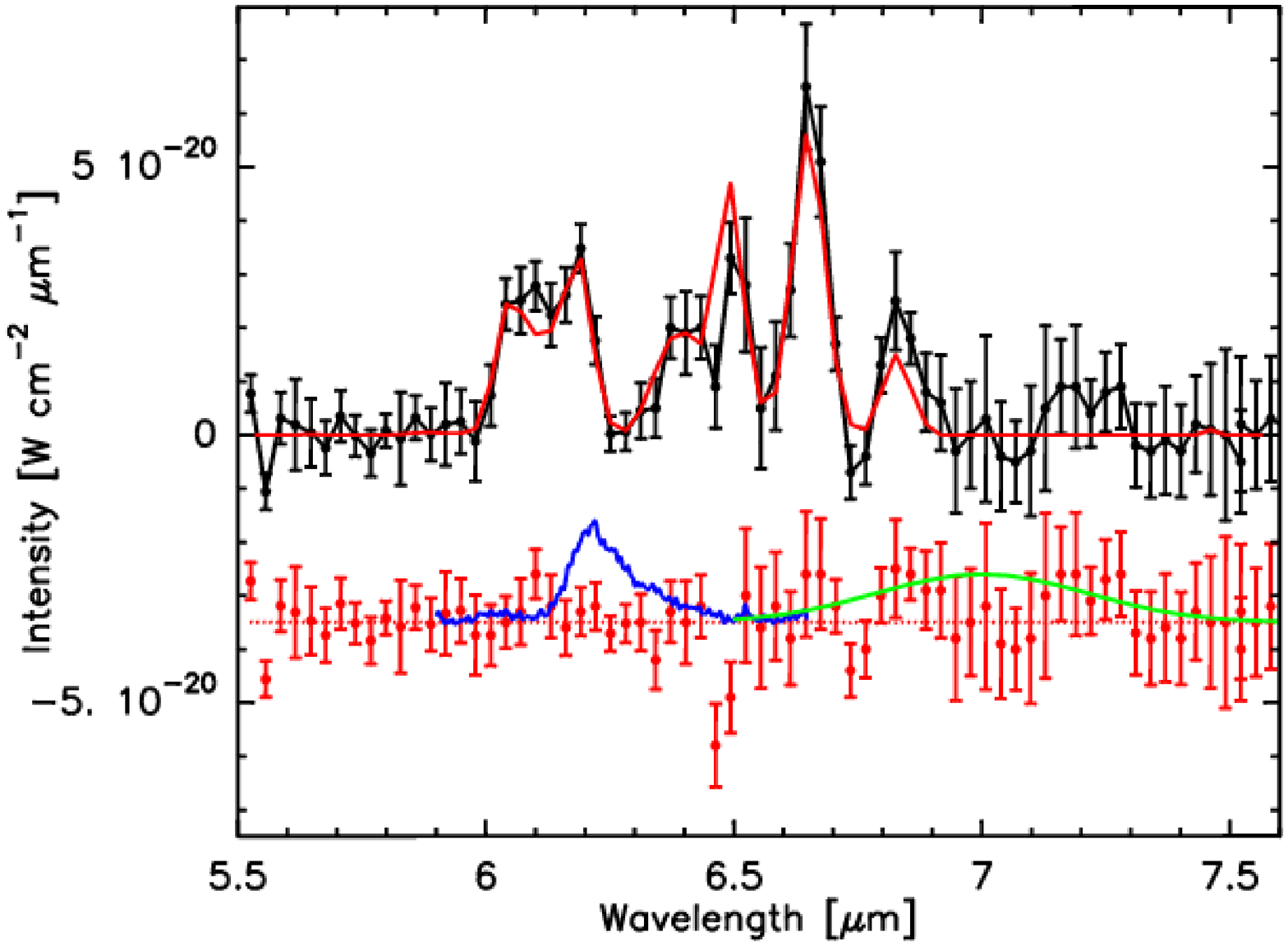}
\end{center}
\caption{Model fit to the 2006 May 27.56~UT (pre-perihelion) spectra of
comet 71P/Clark with extraction apertures
of 3.7\arcsec $\times$ 5.6\arcsec~(top) and 3.7\arcsec $\times$
17.9\arcsec~(bottom).  Data are shown in black dots with errorbars. The
synthetic water spectrum obtained from model fitting with the
ortho-to-para ratio set to OPR = 3 is shown in red. The retrieved
$T_{rot}$ is $14\pm4$~K and $5\pm13$~K for the small and large
apertures, respectively. The residual spectrum is shown in red on the
bottom, on which are superimposed in arbitrary units an interstellar PAH
spectrum typical of class A sources from \citet{pee02} ({\it blue}
spectrum), and a model of carbonate emission from \citet{lisse06} ({\it
green} spectrum).
\label{fig:71pfit}}
\end{figure}

In the case of 71P/Clark, model fitting was performed for two different
extraction apertures along the slit, 3.7\arcsec $\times$ 5.6\arcsec~(top
of Fig.~\ref{fig:71pfit}) and 3.7\arcsec $\times$ 17.9\arcsec~(bottom of
Fig.~\ref{fig:71pfit}). For the small extraction aperture, there is
significant disagreement at 6.35--6.40~\micron{} (this region was masked
during the fitting procedure to obtain successful results in the other
domains). The intensity of the water band is less intense than expected
from modeling, resulting in negative residuals. This low intensity is
puzzling. At low temperatures, the 6.40~\micron{} peak arises mainly from
the $1_{01}-1_{10}$ and $1_{11}-2_{02}$ lines, whereas the 6.50~\micron{}
peak comes mainly from the $1_{01}-2_{12}$ and $2_{12}-2_{21}$ lines. The
ratio of their intensities $I$(6.40 $\mu$m)/$I$(6.50 $\mu$m) decreases
with decreasing $T_{rot}$ \citep[see Fig.~5 of][]{woodw07}, but reaches
a constant value of $\sim$ 0.4--0.5 at low $T$ and fluorescence
equilibrium. For the large extraction aperture, the discrepancy is
shifted towards the 6.45--6.50~\micron{} domain (Fig.~\ref{fig:71pfit}).
Therefore it is likely that the misfit in the 6.35--6.50~\micron{} range
is related to incorrect  background subtraction and/or instrumental
artifacts. Outside this spectral domain, the agreement between our model
of H$_{2}$O emission and the \spitzer{} spectra of 71P/Clark is good.

The best fits to the spectra of 71P/Clark correspond to rotational
temperatures $T_{rot}$ of $5 \pm 13$~K (large extraction aperture) and
$14 \pm 4$~K (small extraction aperture), both consistent with $T_{rot}$
$<$ 18~K (similar values are found for OPR = 2.5 and 3.0). Indeed, the
6.50~\micron{} peak is less intense than the 6.65~\micron{} peak only for
$T_{rot}$ $<$ 20 K \citep[see Fig. 5 of][where the emergent intensity of
the H$_{2}$O $\nu_{2}$ band is plotted as a function of rotational
temperature]{woodw07}. The reverse was observed in \spitzer~spectra of
C/2003 K4 (LINEAR) from which $T_{rot}$ values $\sim$ 30 K were
determined \citep{woodw07}.

\begin{figure}
\begin{center}
\includegraphics[angle=270,scale=.40]{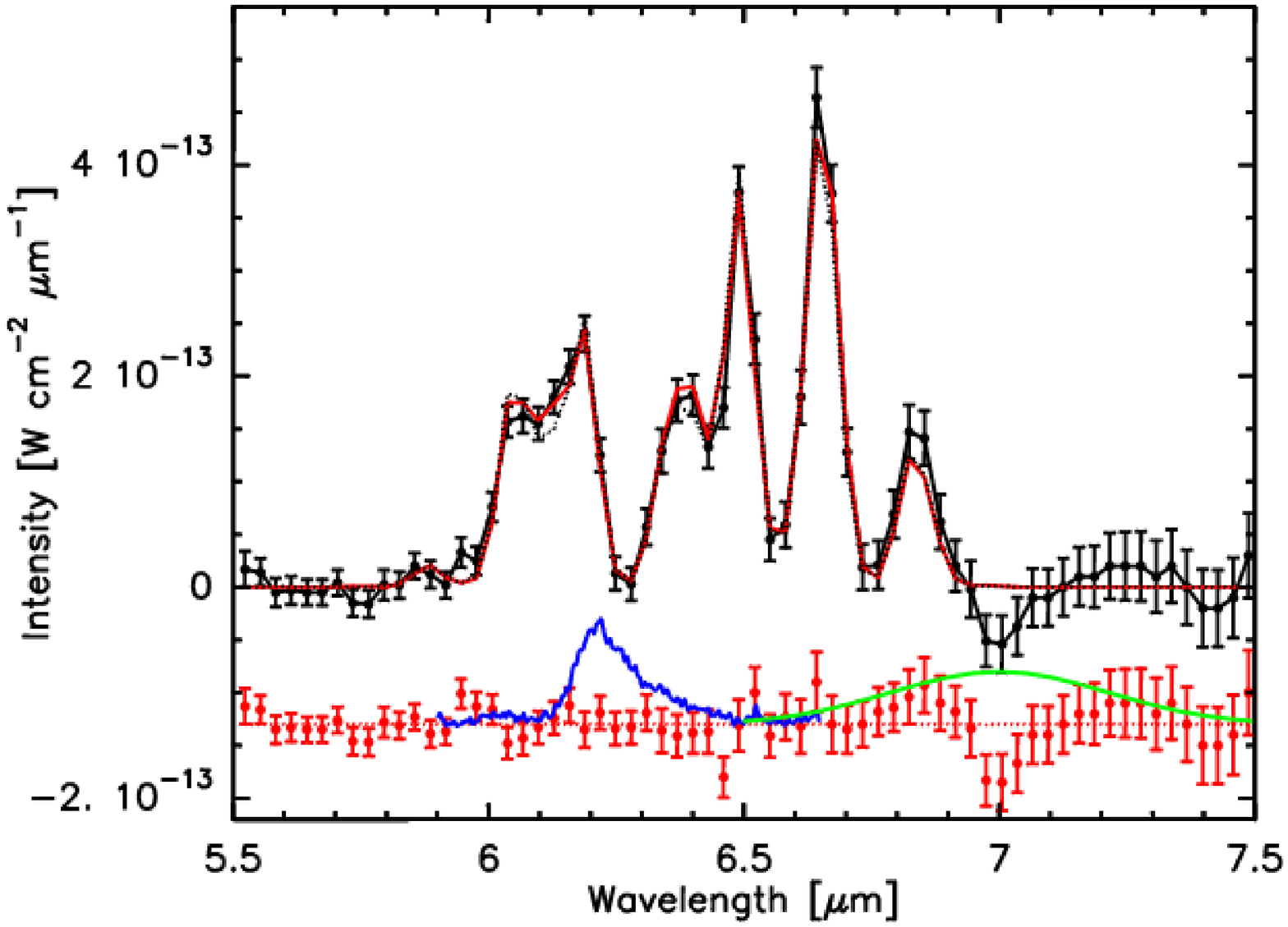}
\includegraphics[angle=270,scale=.40]{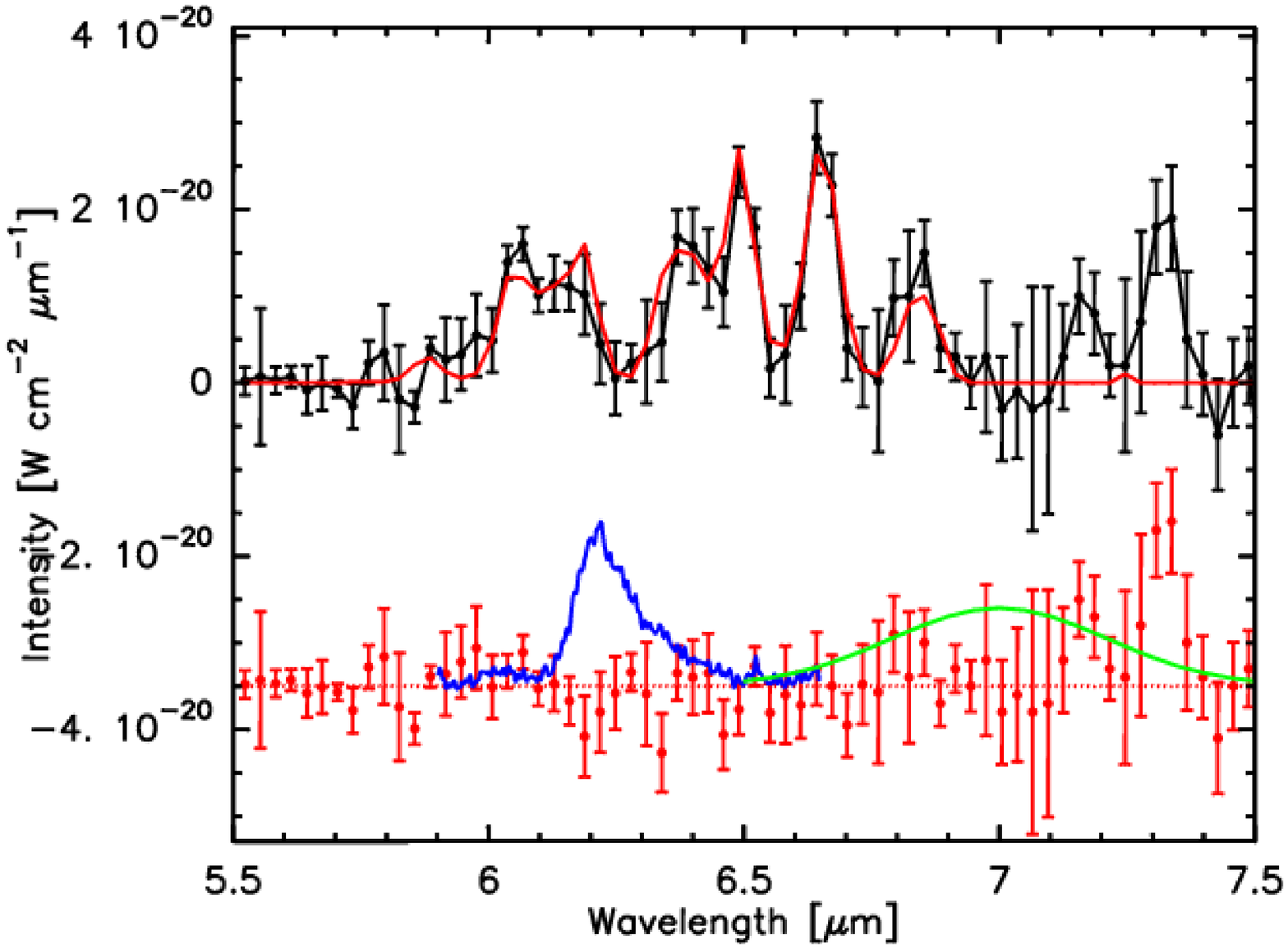}
\end{center}
\caption{Model fits to the 2005 October
27.56~UT (pre-perihelion, top) and 2006 May 16.22~UT
(post-perihelion, bottom) spectra of comet C/2004 B1
(LINEAR) obtained with an extraction aperture of
3.7\arcsec $\times$ 17.9\arcsec.  Data are shown in
black dots with errorbars, while
other colored curves are as described in Fig.~\ref{fig:71pfit}. Model
fitting to the pre-perihelion spectrum yields
$T_{rot} = 13.6\pm2.1$~K and OPR = $2.31 \pm 0.18$. 
The best fit obtained with the OPR fixed to OPR = 3 ($T_{rot} = 13.4$ K)
is shown by  a black dotted line. For the
post-perihelion spectrum, the OPR is fixed to a value of 2.3, and
the derived rotational temperature is $T_{rot} = 23\pm4$~K.
\label{fig:post-2004b1fit}}
\end{figure}

The signal-to-noise ratio achieved in the C/2004 B1 (LINEAR)
pre-perihelion water spectrum ($\sim$50 in-band intensity on the
3.7\arcsec $\times$ 17.9\arcsec~extraction aperture considered for
model fitting) permits a determination of the water ortho-to-para
ratio (OPR), as done from \spitzer{} spectra of C/2003 K4 (LINEAR)
\citep{woodw07}. The best fit, shown in
Figure~\ref{fig:post-2004b1fit}, corresponds to $T_{rot}$ $\simeq$
$14 \pm 2$~K and OPR $= 2.31 \pm 0.18$. Also
shown in Fig.~\ref{fig:post-2004b1fit} is the best fit obtained
with the OPR fixed at a value of 3. The $\chi^2$ is increased by
$\Delta\chi^2$ = 10 with respect to the best fit obtained with OPR
$= 2.31$.

The agreement between our model and the \spitzer{} spectrum is
excellent, except for the structure at 6.85~\micron{} where the
predicted intensity is lower than observed with a 2--$\sigma$
discrepancy. Taking the 6.05~\micron{} peak as a reference since
its intensity is weakly dependent upon $T_{rot}$ and OPR
\citep[cf., Figs. 5 and 6 of][]{woodw07}, the intensity ratio
$I$(6.85~\micron{})/$I$(6.05~\micron{}) slowly increases with
increasing $T_{rot}$ (and decreasing OPR) \citep[Figs. 5 and 6
of][]{woodw07}. The observed ratio would be typical of $T_{rot}$
$\sim$ 50 K. However, the relative intensities of the other peaks
(in particular, $I$(6.65~\micron{})/$I$(6.05~\micron{}) and
$I$(6.18~\micron{})/$I$(6.05~\micron{})) are indicative of a low
$T_{rot}$. Model fitting to the 5.9--6.7~\micron{} part of the
spectrum yields the same results as those obtained from the whole
spectrum. As discussed in \citet{woodw07}, the band regions most
sensitive to the OPR lie at 6.12 and 6.4~\micron{} where the
emission is dominated by para lines: model fitting with OPR = 3
results in excess emissions centered at 6.12 ~\micron{} and
6.4~\micron{} in the residuals. The discrepancy at 6.85~\micron{}
is likely related to instrumental effects or flaws in the
background subtraction (the negative feature at 7~\micron{} in the
residuals is not explained either).

The best-fit modeled spectrum for C/2004 B1 (LINEAR)
post-perihelion is shown in Figure~\ref{fig:post-2004b1fit}. For
this model fitting, we fixed the ortho-to-para ratio to the value
OPR = 2.3 derived from the pre-perihelion data. The derived
rotational temperature is $T_{rot} = 23 \pm 4$~K. A similar value
($T_{rot}$ = 20~K) is obtained with OPR = 3 (fit shown in
Fig.~\ref{fig:spectra}) and OPR = 2.5 ($T_{rot}$ = 22~K). The agreement
between our model and the \spitzer{} spectrum is rather good. The
residuals with respect to the observed spectrum do not show any
excess emission in the wavelength range 5.8--7.1~\micron{} of
water $\nu_{2}$ band emission.

\subsection{\textit{Water Production Rates}}
\label{sec:qh2o}

From the intensity of the $\nu_{2}$ H$_{2}$O band measured on the
spectra, we computed the water production rates of 71P/Clark and C/2004
B1 (LINEAR) using a Haser model for the water density (Table~\ref{table:fit}).
Because the observations of comet 71P/Clark were affected by
pointing offsets due to uncertainties in the ephemeris
(\S~\ref{sec:obs}), the production rates retrieved from the
intensities measured in a small (3.7\arcsec $\times$ 5.6\arcsec)
and large (3.7\arcsec $\times$ 17.9\arcsec) extraction aperture
were used as a complementary check to the pointing offsets.
Consistent values for $Q$(H$_2$O), equal to ($1.08 \pm 0.08$) $\times
10^{28}$~molecules~s$^{-1}$ and ($1.04 \pm 0.08$) $\times
10^{28}$~molecules~s$^{-1}$ for the small and large extraction
apertures respectively, are obtained applying offsets
corresponding to the actual position of the nucleus with respect
to the targeted position. This derived production rate for the
2006 perihelion is consistent with indirect estimations from the
visual brightness at previous perihelia \citep{Sekanina93}.

The water production rate derived for C/2004 B1 (LINEAR) at
$r_h$ = 2.21 AU pre-perihelion is $Q$(H$_2$O) = ($1.03 \pm 0.02$)
$\times 10^{28}$~molecules~s$^{-1}$. At $r_h$ = 2.06 AU post-perihelion,
the measured $Q$(H$_2$O) is ($4.74 \pm 0.42$) $\times
10^{27}$~molecules~s$^{-1}$, which suggests that the comet was more active
pre-perihelion. This later conclusion is consistent with measurements of
the total visual magnitude $m_v$ of the comet. For
2005 October 21 and 2006 May 17, i.e., at dates close to the
\spitzer{} observations, values $m_v$ = 12.6 and 12.4 are reported,
which correspond to heliocentric visual magnitudes of 11.1 and 11.8,
respectively \citep{green05,green06}.

\begin{figure}
\begin{center}
\includegraphics[angle=0,scale=.60]{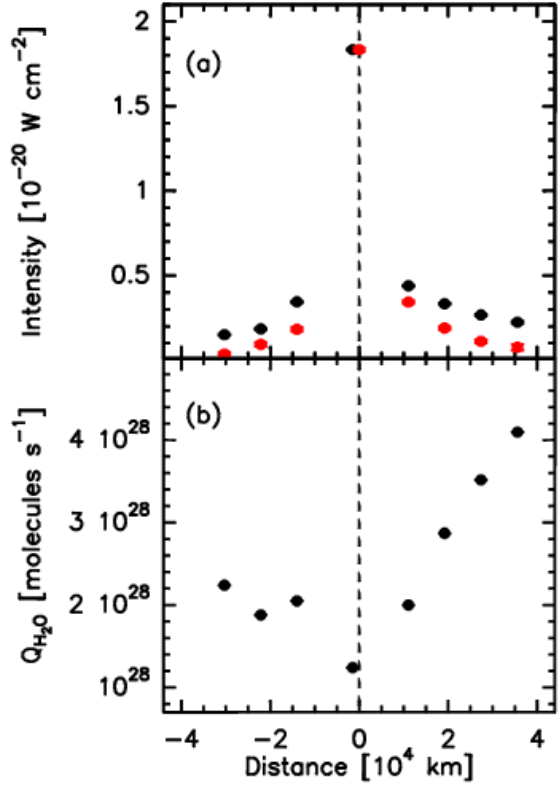}
\end{center}
\vspace{-0.75cm}
\caption{Distribution of water and dust as a function of
offset, $\rho$, with respect
to the nucleus position in C/2004 B1 (LINEAR) at $r_{h}$ = 2.2~AU
(pre-perihelion).  Negative offsets correspond to southern extractions
with respect to the peak in the surface brightness of the coma,
Fig.~\ref{fig:pointing}(a). (a)~Total water $\nu_{2}$ band
intensity between 5.8 and 7~\micron{} (\textit{black circles}) and
dust continuum (\textit{red circles}) surface brightness determined
from equivalent narrow-band continuum (divided by a factor of 17)
measured  at $\lambda = 5.45 \pm 0.08$~\micron{}. Extraction
apertures are 5.6\arcsec~$\times$~5.6\arcsec~, except for the most central
extraction which has an 5.6\arcsec~$\times$~11.1\arcsec{} aperture.
For clarity, the central (offset $= -1$\arcsec) dust point has been
shifted to an offset equal to zero.
The water intensity varies as $\rho^{-1.13}$ for the southern extractions,
and as $\rho^{-0.54}$ for the northern extractions, while the dust
emission (average of the northern and southern coma values) decreases
with increasing offset as $\rho^{-1.30}$. (b)~The $Q$-curve, i.e.,
the apparent $Q$(H$_2$O) production rates required to explain the
H$_{2}$O band intensity at the different offsets in the assumption of a
Haser parent molecule distribution. \label{fig:B1-distribution}}
\end{figure}

Water production rates were computed assuming a Haser parent-molecule
distribution for the water density with the parameters given
in \S~\ref{sec:h2ofit}. The signal-to-noise ratio achieved
for C/2004 B1 (LINEAR) pre-perihelion
permits the investigation of the radial distribution of water
throughout the IRS SL2 spectral map (Fig.\ref{fig:pointing}a). The
intensity of the $\nu_{2}$ band along nine extractions is plotted as a
function of offset $\rho$ in Fig.~\ref{fig:B1-distribution}(a). The
$\nu_{2}$ intensity varies as $\rho^{-0.54}$ north of the nucleus which
suggests that the water distribution is extended in this
direction. South of the nucleus, the
intensity varies as $\rho^{-1.13}$, which is consistent with a Haser
distribution. For comparison (Fig.~\ref{fig:B1-distribution}a),
the surface brightness of the dust emission
(derived from continuum measurements at 5.45~$\pm$~0.08~\micron{}
in each of the nine apertures) decreases with
increasing offset as $\rho^{-1.30}$ (southern and northern coma averaged).

The extended water distribution is best evidenced in the
``$Q$-curve'' \citep[cf.,][]{mumma03} shown in
Fig.~\ref{fig:B1-distribution}(b) which plots
the water production required to explain the H$_{2}$O band intensity
at the different offsets, assuming a Haser distribution. The
apparent $Q$(H$_2$O) increases by more than a factor of 2
from 1 $\times$ 10$^4$~km to 4 $\times$ 10$^4$~km. This extended
distribution is suggestive of the presence of subliming icy
grains in the coma of C/2004 B1 (LINEAR). At 2.1~AU from the Sun the
lifetime of dirty ice grains ranges from $10^{4} - 10^{6}$~s for
grains of radii $10^{-2} - 1$~cm \citep{beer06}. This
lifetime range corresponds to displacements within the coma of
typically $10^{3} - 10^{4}$~km prior to complete grain sublimation.
The lifetime of pure ice grains is considerably larger
\citep{beer06}. Our measurements are most easily
explained if pure icy grains or ices mixed with poorly
absorbing materials are present in the coma of comet C/2004 B1 (LINEAR).
However, modeling of such sublimation processes is beyond the
scope of this paper.

\subsection{\textit{Rotational Temperatures}}

The measured rotational temperatures (Table~\ref{table:fit}) may
not correspond to physical gas temperatures. As the water
molecules expand in the coma, the rotational populations within
the ground vibrational state relax to a cold fluorescence
equilibrium. Thermal equilibrium breaks down at distances which
depend on the density of the collisional partners (mainly H$_{2}$O
molecules and electrons), and which are directly related to the
water production rate. Given the low water production rate
of 71P/Clark and C/2004 B1 (LINEAR), and the observational
circumstances (field of view, slit offset), we may expect
molecules with non-LTE rotational populations within \spitzer{} IRS
SL2 slit.  Using a model of water rotational excitation and the
methodology explained in \citet{woodw07}, we have computed
synthetic $\nu_{2}$ band profiles for the observing conditions of
71P/Clark and C/2004 B1 (LINEAR) from \spitzer{} and the
field-of-view corresponding to the extraction apertures. Effective
$T_{rot}$ were derived by fitting the synthetic spectra with the
same procedure as used for the observed spectra.

For 71P/Clark, the predicted effective $T_{rot}$ at 8\arcsec~from the
nucleus is 11--12~K for a kinetic temperature $T_{kin}$ in the range
10--50~K and $x_{ne}$ = 1, where
$x_{ne}$ is a multiplying factor to the electron density
normalized to the Giotto 1P/Halley measurements. If our observations
had been executed with the slit centered on the nucleus,
we would have expected $T_{rot}$ = 25 to 29~K with these model
parameters. These results are for an aperture of 3.7\arcsec $\times$
17.9\arcsec, but similar numbers are obtained for a 3.7\arcsec $\times$
5.6\arcsec~aperture: 31--39 K on nucleus, 12--13~K for an offset of 8\arcsec.
The size of the collisional region where
water molecules are thermalized to the gas kinetic temperature is small
with respect to the field of view, hence the predicted rotational
temperature is weakly sensitive to the assumed $T_{kin}$. A relatively
high $T_{rot}$ is expected for a slit centered on the nucleus 
because molecules excited by collisions
with hot electrons have a significant contribution to the signal.
Rotational temperatures measured in 71P/Clark (14 $\pm$ 4~K and 5 $\pm$
13~K for the small and large apertures, respectively) are consistent
with modeling.

Model calculations pertaining to the observational circumstances of
the C/2004 B1 (LINEAR) pre-perihelion data predict $T_{rot}$ = 17--19~K
for a 3.7\arcsec $\times$ 17.9\arcsec~aperture and kinetic temperatures
of 10 to 50 K. Note that low gas temperatures may be expected for this
weakly active (\S~\ref{sec:qh2o}) comet observed at 2.1--2.2 AU from the
Sun. Overall, the agreement with the measured  $T_{rot}$ value
(13.6 $\pm$ 2.1 K) is satisfactory.

In the case of C/2004 B1 (LINEAR) post-perihelion,
the expected $T_{rot}$ for a 3.7\arcsec $\times$ 17.9\arcsec~extraction
centered on the nucleus is $T_{rot}$ = 17~K assuming $T_{kin} = 30$~K,
with little dependency on the assumed $T_{kin}$ since water molecules
with non-LTE rotational populations are sampled. The measured $T_{rot}$
of 23 $\pm$ 4~K is consistent with the model within 2-$\sigma$. A better
agreement could be obtained by increasing the electron density through the
$x_{ne} = 1$ parameter.

\subsection{\textit{Ortho-to-para Ratio in C/2004 B1 (LINEAR)}}

The ortho-to-para ratio of $2.31  \pm 0.18$  measured in C/2004 B1 (LINEAR)
corresponds to a spin temperature $T_{spin}$ = $26^{+3}_{-2}$~K.
The ortho-to-para ratio in water has been measured in a dozen of
comets \citep[see][for a compilation of measurements before 2007]{bbonev07},
and ranges from $\sim$ 1.8 to the equilibrated value of 3 ($T_{spin}$
from $\sim$ 23 to $>$ 40 K). The OPR measured in C/2004~B1 (LINEAR) is
in the low range of measured values in Oort cloud comets, yet comparable
to OPR values reported by \citet{mumma88} and \citet{dello05} for comets
1P/Halley and C/2001 A2 (LINEAR). Our new measurement for
C/2004~B1 (LINEAR) suggests that OPR variation clearly exists 
among the population of Oort cloud comets.

The meaning of nuclear spin temperatures in cometary molecules is
puzzling \citep{kawa04,cro07}. Nuclear spin temperatures may be
related to the thermal conditions experienced by the molecules
since their incorporation into cometary nuclei, or connected to
their formation conditions and subsequent OPR reequilibration
in the solar nebula. \citet{kawa04} have shown that there is no clear
trend between the nuclear spin temperatures, production rate, heliocentric
distance and orbital period of the comet, suggesting that the nuclear
spin temperature is pristine \citep{kawa04}. The measurement of the
OPR (and corresponding nuclear spin temperature) in C/2004 B1 (LINEAR) is
derived from observations conducted when the comet was at a heliocentric
distance, $r_h$ = 2.21 AU. This heliocentric distance is intermediate between
the range $r_h$ = 0.8--1.2 AU, where most OPR measurements were acquired, and
the measurement in C/1995 O1 (Hale-Bopp) at 2.9 AU from the Sun. Our
\spitzer{} measurement provides a new data point for future investigations
of $T_{spin}$ variations with heliocentric distance based on a statistically
significant data set.

\subsection{\textit{Residual Emissions}}
\label{sec:residuals}

\begin{figure}
\begin{center}
\includegraphics[angle=0,scale=.90]{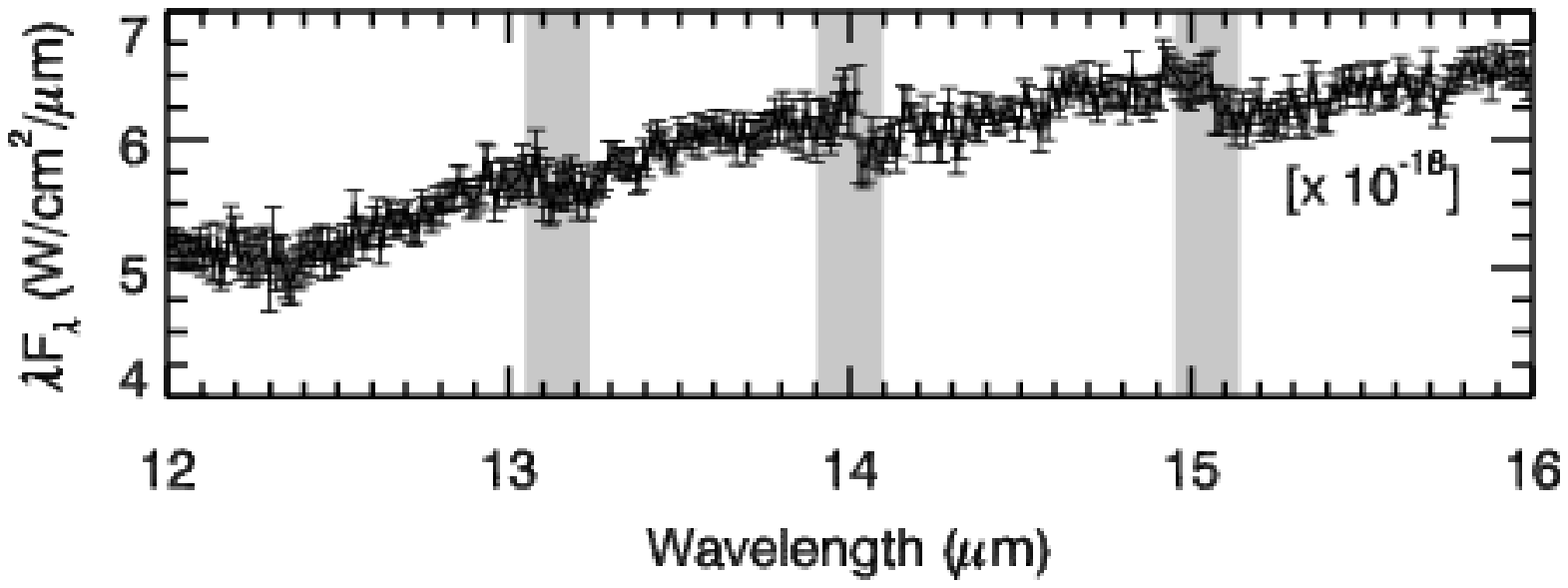}
\end{center}
\caption{The \spitzer{} IRS short-high (SH) spectra of 
comet C/2004 B1 (LINEAR)
over the wavelength range from 12.0~\micron{} to 16.0~\micron{} where
narrow resonances from the out-of-plane and in-plane bending mode
of the (CO$_{3}$)$^{2-}$ ion in carbonate species are expected
\citep{kemper02}. The grey vertically shaded areas indicate
the wavelength locations of the spectral order overlap in the
instrument and the discontinuity of the spectra in these regions
are artifacts. The IRS spectra are featureless in the wavelength
regions of the carbonate resonances (narrow features near 12.7~\micron{}
and 14.8~\micron), suggesting that carbonates are not present with
large abundance, if at all, in the emitting coma materials.
\label{fig:carbonates}}
\end{figure}

The modeled water spectrum wholly accounts for the emission in
excess of the dust continuum background in the two comets
(Figs.~\ref{fig:71pfit} and~\ref{fig:post-2004b1fit}). There is
no emission excess at 6.2~\micron{} that could be attributed to
PAH emission. There is not significant evidence for moderately
broad ($\Delta\lambda \simeq 0.5$~\micron) emission peaking at
7.0~\micron{} (arising from asymmetric stretching vibration of
C--O) that could be due to carbonates minerals \citep{kemper02,
cro08}. Such emission, if present, would have made the
6.85~\micron{} peak of the H$_{2}$O band of less contrast than
observed. In addition if carbonates where present, other narrow
($\Delta\lambda \ltsimeq 0.2$~\micron) emission features near
12.7~\micron{} and 14.8~\micron{} should be present
\citep{kemper02}. The IRS spectra of comets C/2004~B1 (LINEAR) and
71P/Clark over this wavelength interval are featureless,
Fig.~\ref{fig:carbonates}. Thus, carbonates are not present (or
significantly abundant) in the comet comae created through
quiescent sublimation processes or dermal jets from isolated
active areas on the nucleus based on analysis of \spitzer{}
spectra of three comets.

Excess emission is present at 7.1--7.3~\micron{} in the spectrum
of 71P/Clark, and between 7.15 and 7.45~\micron, peaking at
7.32~\micron{} in the post-perihelion spectrum of C/2004 B1
(LINEAR) (Figs.~\ref{fig:71pfit} and ~\ref{fig:post-2004b1fit}).
Excess emission near 7.3~\micron{} was also observed in the
\spitzer{} spectrum of C/2003 K4 (LINEAR) \citep{woodw07}. The
emission feature observed in C/2004 B1 (LINEAR) post-perihelion closely
resembles that observed for C/2003 K4 (LINEAR), but is much more
intense relative to the intensity of the water band than in C/2003
K4 (LINEAR): $\sim$ 10\% the intensity of the water band in the
post-perihelion spectrum of C/2004 B1 (LINEAR),
$\sim$ 2\% in C/2003 K4 (LINEAR) \citep{woodw07}. Its
contrast with respect to the continuum background is also more
important in C/2004 B1 (LINEAR). The feature is not present
in the pre-perihelion spectrum of  C/2004 B1 (LINEAR). This suggests 
that the feature is not related to the absolute brightness of the comet. As
discussed in \citet{woodw07}, the origin of this feature is
unclear. Based on examination of spectral extractions of standard
stars used to calibrate the IRS campaign, this feature is not
likely an instrumental artifact, although we cannot completely
exclude this possibility.

\section{CONCLUSION}
\label{sec:conclusion}

We have presented \textit{Spitzer} IRS spectra of the water
$\nu_{2}$ fundamental vibrational band near 6.5~\micron{} in the
Jupiter-family comet 71P/Clark and the dynamically new comet
C/2004~B1 (LINEAR). Modeling of the IR spectra using a
fluorescence model of the water emission yielded rotational
temperature and production rate measurements. The measured low
rotational temperatures are consistent with water excitation models
which predict that water molecules within the \spitzer{} IRS slit
have partially relaxed (non-LTE) rotational populations. The water
ortho-to-para ratio (OPR) was constrained to $2.31 \pm 0.18$ in
C/2004~B1 (LINEAR) from its pre-perihelion spectrum observed with a
high signal-to-noise ratio. The corresponding spin temperature
$T_{spin}$ =  $26^{+3}_{-2}$~K is in the low range of
measured values in other Oort cloud comets.  In addition,
the spatial distributon of water observed (pre-perihelion) in the
northern coma of C/2004 B1 (LINEAR) is more extended than that
expected for nucleus outgassing which suggests the presence of
subliming icy grains.

Although the 5.8--7.6~\micron{} spectral domain is rich in spectral
signatures of gaseous compounds and minerals our spectra show no
incontrovertible evidence for carbonates, or PAH emission.  Thus,
reported claims that these astrobiologically significant species
are present \citep[e.g.,][]{lisse07, lisse06} remain controversial
\citep[i.e.,][]{cro08, dbm_dps07, woodw07}. As seen in comet
C/2003~K4 (LINEAR) \citep{woodw07}, residual emission at
7.1--7.3~\micron{} is present in spectra of 71P/Clark and
C/2004~B1 (LINEAR) whose origin is unclear.

\acknowledgements

This work is based on observations made with the {\it Spitzer}
Space Telescope, which is operated by the Jet Propulsion
Laboratory, California Institute of Technology under a contract
with NASA. Support for this work was provided by NASA through an
award issued by JPL/Caltech.  Support for this work also was
provided by NASA through contracts 1263741, 1256406, and 1215746
issued by JPL/Caltech to the University of Minnesota.  C.E.W. also
acknowledges support from the National Science Foundation grant
AST-0706980. We thank the anonymous referee for helpful
comments, and N. Biver, D.~E. Harker and D. Pelat for enlightening 
discussions.

{\it Facilities:} \facility{Spitzer (IRS)}

\clearpage


\clearpage

\begin{deluxetable}{lccccccccc}
\tabletypesize{\footnotesize}
\tablewidth{0pt}
\rotate
\tablecaption{WATER IN COMETS: MODEL FITS AND PRODUCTION
RATES\label{table:fit}}
\tablecolumns{10}
\tighten
\tablehead{
\colhead{Obs Date} &
\colhead{$\alpha$\tablenotemark{a}} &
\colhead{$r_h$} &
\colhead{$\Delta_{Spitzer}$\tablenotemark{b}} &
\colhead{Extraction} &
\colhead{Band Intensity\tablenotemark{c}} &
\colhead{$T_{rot}$} &
\colhead{OPR} &
\colhead{($\chi^2_\nu$)\tablenotemark{d}} &
\colhead{$Q_{\rm H_{2}O}$}
\\
 \colhead{(UT)} &
 \colhead{(deg)} &
 \colhead{(AU)} &
 \colhead{(AU)} &
 \colhead{(\arcsec~$\times$~\arcsec)} &
 \colhead{(10$^{-21}$ W~cm$^{-2}$)} &
 \colhead{($K$)} &
 \colhead{} &
 \colhead{} &
 \colhead{(10$^{27}$ molecules~s$^{-1}$)}
}
\startdata
\underbar{71P/Clark}\\
2006 May 27.56 &37.8 &1.57 & 0.92 & 3.7 $\times$ 17.9\tablenotemark{e} &
 \phantom{1}17.0 $\pm$ 1.29 & \phantom{0}5 $\pm$ 13 & 3.0\tablenotemark{f} & 
0.90 & 10.43 $\pm$ 0.79\\
\\

\underbar{C/2004 B1 (LINEAR)}\\
2005 Oct 15.22 &27.5 & 2.21 & 2.03 & 3.7 $\times$ 17.9 &
13.74 $\pm$ 0.28 & 13.6 $\pm$ 2.1 & 2.31 $\pm$ 0.18 & 0.97 & 10.32 $\pm$ 0.21\\

2006 May 16.22 &28.7 & 2.06 & 1.61 & 3.7 $\times$ 17.9 &
 \phantom{1}9.20 $\pm$ 0.82 & 23 $\pm$ 4 & 2.3\tablenotemark{f} & 
0.90 & \phantom{1}4.74 $\pm$ 0.42\\
\enddata
\tablenotetext{a}{Phase angle (defined as vertex angle of
Sun-Comet-\spitzer).}
\tablenotetext{b}{Distance to \spitzer{} spacecraft.}
\tablenotetext{c}{Band intensity between 5.8~\micron{} and 7~\micron.}
\tablenotetext{d}{Reduced $\chi^2$ between 5.85~\micron{} and 7~\micron{}. }
\tablenotetext{e}{Offset with respect to the
nucleus position by $\sim$ 8\arcsec{} (see \S~\ref{sec:obs}).}
\tablenotetext{f}{Assumed.}
\end{deluxetable}

\clearpage




\clearpage


\clearpage

\clearpage


\clearpage


\clearpage


\clearpage


\clearpage
\end{document}